\documentclass[aip,jcp,reprint,twocolumn,amsmath,amssymb,amsfonts,floatfix,letterpaper,longbibliography]{revtex4-2}
\usepackage{mathptmx}
\usepackage[scaled=0.90]{helvet} 
\usepackage{courier} 
\usepackage{bm} 
\newcommand{\symbf}[1]{\boldsymbol{#1}}
\newcommand{\symbfup}[1]{\mathbf{#1}}
\usepackage{hyperref}
\hypersetup{colorlinks=true, citecolor=blue, urlcolor=blue, linkcolor=black}

\usepackage{xkeyval}
\makeatletter
\newcommand*\acs@warning{\PackageWarning{achemso}}
\newcommand*\acs@ifundefined[1]{
  \begingroup\expandafter\expandafter\expandafter\endgroup
  \expandafter\ifx\csname #1\endcsname\relax
    \expandafter\@firstoftwo
  \else
    \expandafter\@secondoftwo
  \fi
}
\RequirePackage{xkeyval}
\newcommand*\acs@keyval@bool[2]{
  \acs@ifundefined{acs@#1#2}{
    \acs@warning{Unknown option `#2' for key #1}
  }{
    \@nameuse{acs@#1#2}
  }
}
\define@key{acs}{usetitle}[true]{
  \acs@keyval@bool{articletitle}{#1}
}
\makeatother
\setkeys{acs}{usetitle = true}

\usepackage{xparse}
\ExplSyntaxOn
\NewDocumentCommand{\mref}{m}{\quinn_mref:n {#1}}
\seq_new:N \l_quinn_mref_seq
\cs_new:Npn \quinn_mref:n #1
 {
  \seq_set_split:Nnn \l_quinn_mref_seq { , } { #1 }
  \seq_pop_right:NN \l_quinn_mref_seq \l_tmpa_tl
  ( 
  \seq_map_inline:Nn \l_quinn_mref_seq
    { \ref{##1},\nobreakspace } 
  \exp_args:NV \ref \l_tmpa_tl 
  ) 
 }
\ExplSyntaxOff

\usepackage{graphicx}
\usepackage{braket}
\usepackage{mleftright}
\usepackage[usenames,dvipsnames]{color}
\AtBeginDocument{\usepackage{booktabs}}
\usepackage{multirow}
\usepackage[version=4]{mhchem} 
\usepackage{adjustbox} 

\usepackage{dcolumn} 
\usepackage{cancel} 

\usepackage{titlesec}
\usepackage{mathtools} 

\definecolor{LightGrey}{rgb}{0.85,0.85,0.85}

\definecolor{DarkGrey}{rgb}{0.2,0.2,0.2}
\definecolor{DarkRed2}{rgb}{0.9,0,0.00}
\definecolor{DarkGreen}{rgb}{0.0,0.3,0.0}
\definecolor{Green50}{rgb}{0.0,0.5,0.0}
\definecolor{DarkYellow}{rgb}{0.5,0.42,0.0}
\definecolor{DarkRed}{rgb}{0.5,0.0,0.0}
\definecolor{DarkRed3}{rgb}{0.3,0.0,0.0}
\definecolor{TextHighlight}{rgb}{0.4,0,0.00}
\definecolor{MathHighlight}{rgb}{0.7,0,0.00}
\definecolor{TextLowlight}{rgb}{0.4,0.4,0.4}
\definecolor{CommentGray}{rgb}{0.3,0.3,0.3}

\usepackage{listings}

\definecolor{gColor044}{rgb}{0.0,0.4,0.4}
\definecolor{gColor024}{rgb}{0.0,0.2,0.4}
\definecolor{gColor026}{rgb}{0.0,0.2,0.6}
\definecolor{gColor048}{rgb}{0.0,0.4,0.8}
\definecolor{gColor408}{rgb}{0.4,0.0,0.8}
\definecolor{gColor480}{rgb}{0.4,0.8,0.0}
\definecolor{gColor084}{rgb}{0.0,0.8,0.4}
\definecolor{gColor840}{rgb}{0.8,0.4,0.0}
\definecolor{gColor804}{rgb}{0.8,0.0,0.4}
\definecolor{gColor444}{rgb}{0.4,0.4,0.4}

\definecolor{cdcolor}{rgb}{.6,.6,.6}

\definecolor{mypurple}{rgb}{.8,.0,.4}

\definecolor{myotherhl}{rgb}{0.,.4,.8}

\definecolor{myotherhll}{rgb}{0.,.5,.2}

\newcommand{\figref}[1]{Fig.~\ref{#1}}

\newcommand{\secref}[1]{Sec.~\ref{#1}}
\newcommand{\appxref}[1]{Appx.~\ref{#1}} 

\newcommand{\Eqref}[1]{Eq.~\eqref{#1}} 
\newcommand{\Eqsref}[1]{Eqs.~\mref{#1}} 

\newcommand{\fn}[2]{\mathinner{#1(#2)}}
\newcommand{\Fn}[2]{\mathop{#1}\mleft(#2\mright)}

\newcommand{\InRange}[3]{\allowbreak{#1=\allowbreak#2\allowbreak\mathellipsis\allowbreak#3}}

\newcommand{\IndexedSetI}[2]{\{#1\}_{#2}} 
\newcommand{\IndexedSetE}[4]{\{#1\}_{#2=#3}^{#4}} 

\newcommand{\progname}[1]{{\texttt{#1}}}

\newcommand{\emb}[1]{\mbox{$#1$}} 

\renewcommand{\d}{\mathord{\mathrm{d}}}

\newcommand{\efrac}[2]{\emb{\frac{#1}{#2}}} 

\renewcommand{\vec}[1]{\symbf{#1}}
\newcommand{\opvec}[1]{\smash{\hat{\symbf{#1}}}}
\newcommand{\mat}[1]{\symbfup{#1}}

\newcommand{\norm}[1]{\left\Vert #1 \right\Vert}
\newcommand{\abs}[1]{\left| #1 \right|}

\newcommand{\Exp}[1]{\Fn{\exp}{#1}}

\newcommand{\lin}{\operatorname{span}}

\renewcommand{\phi}{\varphi}
\renewcommand{\epsilon}{\varepsilon}

\newcommand{\Nnuc}{\smash{N_\mathrm{nuc}}}

\newcommand{\subpoint}[1]{\medskip\noindent\textsf{#1}.\;}

\definecolor{oldterm}{rgb}{0.4,0.0,1.0}
\definecolor{newterm}{rgb}{1.0,0.5,0.0}

\DeclareMathOperator*{\argmax}{arg\,max}

\newcommand{\Nelec}{N}
\newcommand{\Nloc}{{N_\mathrm{orb}}} 
\newcommand{\norb}{\Nloc}
\newcommand{\lmi}[1]{\phi_{#1}} 
\newcommand{\lmo}[1]{\phi'_{#1}} 
\newcommand{\denslmo}[1]{\rho'_{#1}} 
\newcommand{\lki}[1]{\ket{\lmi{#1}}} 
\newcommand{\lko}[1]{\ket{\lmo{#1}}} 

\newcommand{\vN}{\mathrm{vN}}
\newcommand{\ER}{\mathrm{ER}}
\newcommand{\FB}{\mathrm{FB}}
\newcommand{\PM}{\mathrm{PM}}
\newcommand{\gPM}{\mathrm{genPM}}

\newcommand{\Nlop}{N_{\smash{\mathrm{op}}}}
\newcommand{\lop}[1]{\hat o_{#1}} 
\newcommand{\ilop}{F}
\newcommand{\LopMel}[3]{o_{#1#2}^{#3}} 
\newcommand{\npl}{\mathrm{NPL}}
\newcommand{\ndl}{\mathrm{NDL}}

\newcommand{\nplFullName}{Nuclear Potential Localization}
\newcommand{\ndlFullName}{Nuclear Distance Localization}
\newcommand{\textnpl}{NPL}
\newcommand{\textndl}{NDL}

\begin{document}
\title{\color{blue}A novel class of $\sigma$/$\pi$-separating orbital localization functionals\vspace*{0.18cm}}

\date{\today}
\author{Gerald Knizia and Elvira Sayfutyarova}
\affiliation{Department of Chemistry, The Pennsylvania State University, University Park PA 16802, USA}

\begin{abstract}
Orbital localization methods have a long history in quantum chemistry, as they provide a convenient way to describe individual chemical bonds and local properties. 
However, only a handful of localization functionals are known up to date, and most localization methods do not separate $\sigma$- and $\pi$-orbitals, which is important to describe chemical bonding patterns and reactivity.
In this work, we present a new class of localization functionals, based on nucleus-electron physical observables, that preserve $\sigma$- and $\pi$-orbitals separation in molecular systems.  
We also present a general mechanism behind $\sigma/\pi$-separating variational localization which can be used for constructing a broader class of $\sigma/\pi$-separating localization functionals from other  local physical quantities.

\end{abstract}

\maketitle

\section{Introduction}\label{sec:Intro}
The ability of electronic structure theory to predict properties of molecular and condensed systems directly from first principles---without any prior knowledge of those systems---has become an indispensible asset in many areas of science.
The vast majority of such calculations are performed with self-consistent field (SCF) methods, such as Kohn-Sham Density Functional Theory (DFT) or Hartree-Fock (HF). In these, the formally \emph{immensely} complex $\Nelec$-electron wave function $\Psi(\vec x_1,\ldots,\vec x_N)$, is approximated by a Slater determinant $\Phi$---a much simpler type of $\Nelec$-electron wave function, built as an anti-symmetrized product of $\Nelec$ one-electron wave functions $\IndexedSetE{\lmi{i}}{i}{1}{\Nelec}$ called $\Phi$'s occupied \emph{molecular orbitals}:
\begin{align}
   \Psi(\vec x_1\ldots\vec x_\Nelec)\approx\Phi(\vec x_1\ldots\vec x_\Nelec) = \frac{1}{\sqrt{\Nelec!}}\det\bigl[\bigl(\lmi{i}({\vec x_j})\bigr)_{i,j=1}^{\Nelec}\bigr].\label{eq:PsiApproxPhi}
\end{align}
As evidenced by the practical success of SCF methods, this approximation is frequently well-warranted; in small benign molecules, it is not uncommon to find $\sqrt{\braket{\Psi|\Phi}}\geq98\%$.
While programs typically compute $\Phi$ in terms of delocalized canonical molecular orbitals 
$\IndexedSetI{\lmi{i}}{i}$, 
which are appropriate for describing certain ionization and excitation processes,\cite{Koopmans:EnergiesOfIndividualElectrons1934,Janak:KsDftEigenvalues1978}
we have previously argued that the analysis of chemical bonding and reactivity is better served by localized intrinsic bond orbitals (IBOs).\cite{knizia:iao,knizia:CurlyArrows,klein:cPCETvsHAT2018}
These are \textit{also} an \emph{exact} representation of $\Phi\approx\Psi$, by virtue of $\Phi$'s invariance to unitary transformations within the occupied orbital space (\emph{vide infra}).

Already in 1930 Fock established the orbital invariance properties of $N$-electron determinants,\cite{fock:UnitaryInvariance} which lie at the heart of orbital localization techniques.
Nevertheless, only a small number of conceptually unique localization approaches have been proposed over time \cite{BenAmor2021}, and
the classical methods of Edminston-Ruedenberg (ER),\cite{edmiston:LocalizedAtomicAndMolecularOrbitals} Foster-Boys (FB),\cite{boys:FosterBoys1,foster:FosterBoys2}  (also known as maximally localized Wannier functions \cite{marzari1997maximally} when applied in condensed systems) and Pipek-Mezey\cite{pipek:PMlocalization} (PM) remain the most prominent to this day.
That being said, these classic methods are not free of problems.
Indeed, 
many modifications have been proposed: some have addressed fundamental flaws of the base method (in particular, the unphysical charge definition in PM\cite{knizia:iao,lehtola:GeneralizedPM}), others have  increased the robustness and domain of applicability,\cite{jonsson:PipekMezeyInSolids2017,clement:PipekMezeyInSolids2021,Senjean:RelativisticIbo2021} and yet others have provided powerful algorithmic improvements.\cite{Lehtola:UnitaryLocalization2013,Luo:VariableMetricLocalization2020}

In the recent past, functional-based orbital localization methods have primarily served as quantitative computational tools; for example, most types of local electron correlation methods in quantum chemistry rely on a localized description of the occupied orbital space.\cite{saebo1985local,saebo1993local, pulay1983localizability, hampel1996local,stoll1996towards,schutz1999low,stoll2005accuracy,neese2009efficient,yang2012orbital}
However, during the last decade, it became widely recognized that functional based orbital localization can serve as a valuable instrument for chemical reasoning, by providing an exact view of quantitative $N$-electron wave functions accessible to human intuition.
This development was spawned by the assertions that IBOs are a legitimate physical manifestation of the empirical Lewis structure,\cite{knizia:iao} and their transformations along a system's reaction coordinate are a direct physical manifestation of the empirical electron flow concept, as encoded by ``arrow pushing'' reaction mechanisms.\cite{knizia:CurlyArrows}

\begin{figure}[h!]
  \centering
  \includegraphics[width=0.9\columnwidth]{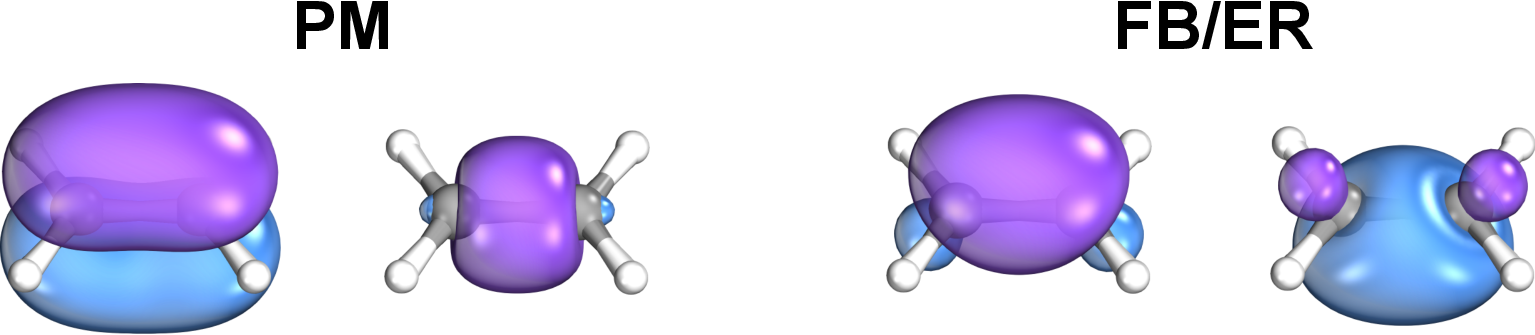}
  \caption{Localized molecular orbitals obtained for ethene with the PM and FB methods.  ER localization yields orbitals which are visually near-indistinguishable from FB orbitals. The orbitals are visualized with an isosurface threshold corresponding to a 80$\%$ density cutoff}
  \label{fig:BentBondsC2H4}
\end{figure}

When used to describe chemical properties and reactivity, molecular $\sigma$- and $\pi$-bonds are fundamentally different.
However, of the classical localization methods, all except for PM (which has a unphysical localization functional) mix the $\sigma$- and $\pi$-manifolds during localization; for example, instead of producing one $\sigma$-bond and one $\pi$-bond in ethene, FB and ER yield two equivalent bent bonds (``banana bonds''; see \figref{fig:BentBondsC2H4}).
In less-obvious bonding scenarios, and in chemical reactions, this behavior can greatly complicate any attempt to interpret those orbitals in chemical terms [for an explicit example, see the Claisen rearrangement in the SI of Ref.~\onlinecite{knizia:CurlyArrows}].

However, apart from the rarely encounterd density-overlap functional of von Niessen,\cite{vonNiessen:OrbitalChargeOverlapLocalization1972}
we are neither aware of any functional-based localization method which fundamentally differs from the three mentioned ones (ER, FB, PM), nor are we aware of any class of localization methods except PM and its generalizations\cite{knizia:iao,lehtola:GeneralizedPM}, which lead to localizations which separate $\sigma$-orbitals and $\pi$-orbitals.
There are also orbital localization approaches which are \emph{not} based on the maximization of a localization functional, but rather involve discrete decisions regarding the most appropriate local representation of a given $N$-electron wave function---most prominently, the methods of the NBO\cite{weinhold:nlmo1985,glendening:NaturalBondOrbitalMethods2012} and AdNDP\cite{zubarev:AdaptiveNaturalDensityPartitioning2008} families fall into this category, but also several entirely disjoint algorithmic approaches.\cite{Aquilante:CholeskyLocalization2006,Zikowski:MinimalMatrixLocalization2009,zhang:SequentialTrafoLocalization2014,damle:QrLocalization2015}
While some of these approaches maintain $\sigma$/$\pi$-separation, we are not aware of such methods, which are based on  manifestly physically observable target criteria;
additionally, the use of discrete decisions in the localization criteria is intrinsically problematic for deriving empirical electron flow pictures of reaction mechanisms by following orbital transformations along a reaction's potential energy surface---something functional based methods \emph{can} achieve.\cite{knizia:CurlyArrows}

In this work, we describe a conceptually new unitary orbital localization procedure---if applied to the occupied orbitals $\IndexedSetI{\lmi{i}}{i}$ of a Slater determinant $\Phi$, the output orbitals $\IndexedSetI{\lmo{i}}{i}$ preserve $\Phi$ exactly (up to global phase), and any physical observable computed from either orbital set will be mathematically identical.
We show evidence that the new method is well suited for the analysis of chemical bonding---in particular, it yields separate $\sigma$- and $\pi$-orbitals, a feature which makes it unique among the previous localization functionals employing only physical obserables.
The resulting orbitals closely resemble IBOs, but are constructed by a simple locality metric entirely formulated in terms of direct physical observables (electron-nuclear distance or electron-nuclear potential energy), 
rather than relying on the abstract auxiliary concept of orbital atomic partial charge.
As practical advantages over IBOs (and other generalized PM-type methods), the new localization reduces orthogonality tails and avoids mixing orbitals with exactly degenerate atomic charge distributions.

\section{Methods}
The new localization methods, as well as all three of the classical localizations, are formulated as a problem of continuous optimization of a \emph{localization functional}.

Concretely, let $\IndexedSetE{\lmi{i}}{i}{1}{\Nloc}$ denote the set of input molecular orbitals,
$\hat U$ a unitary transformation within $\lin\IndexedSetE{\lmi{i}}{i}{1}{\norb}$,
and $\IndexedSetE{\hat U \lmi{i}}{i}{1}{\Nloc}$ the output orbitals:
\begin{align}
   \lko{i} &:= \mathinner{\hat U}\lki{i} := \sum_{k=1}^{\Nloc} \lki{k} U_{ki}. &&(\InRange{i}{1}{\norb}). \label{eq:UnitaryOrbitalTrafo}
\end{align}

Then the localized orbitals are defined by that unitary $\hat U$ which maximizes a method-specific scalar measure of orbital locality, as characterized by a functional $L[\IndexedSetI{\phi_i}{i}]$:
\begin{align}
   \hat U = \argmax_{\text{unitary } \hat U} L\Bigl[\IndexedSetE{\hat U \lmi{i}}{i}{1}{\Nloc}\Bigr].
   \label{eq:UnitaryArgmax}
\end{align}

On the conceptual side, the core distinction between different localization methods is \emph{how} the measure of locality is defined.

For example, the closely related FB and ER functionals both define an orbital $\lmo{i}$'s locality in terms of its spatial electron density distribution 
\begin{align}
   \denslmo{i}(\vec r)=\abs{\smash{\lmo{i}(\vec r)}}^2=\bigl(\lmo{i}(\vec r)\bigr){}^*\lmo{i}(\vec r).\label{eq:OrbitalElectronDensity}
\end{align}

Concretely, FB characterizes \emph{increasing} locality by the density distribution's \emph{decreasing} squared spatial extent:
\begin{align}
   L_{\FB}\Bigl[\IndexedSetI{\lmo{i}}{i}\Bigr] :=& -\sum_{i=1}^{\norb} \iint \denslmo{i}(\vec r_1) \norm{\vec r_1-\vec r_2}^2 \denslmo{i}(\vec r_2)\,\d^3 r_1\d^3 r_2 
\notag\\   
   =& -\sum_{i=1}^{\norb} \braket{\lmo{i}\lmo{i}|\norm{\opvec r_1-\opvec r_2}^2|\lmo{i}\lmo{i}}
   \label{eq:LocFnBoys1}
\end{align}

and ER by its \emph{increasing} Coulomb self-repulsion (ER):
\begin{align}
   L_{\ER}\Bigl[\IndexedSetI{\lmo{i}}{i}\Bigr] :=& \sum_{i=1}^{\norb} \iint \frac{\denslmo{i}(\vec r_1)\denslmo{i}(\vec r_2)}{\norm{\vec r_1-\vec r_2}}\,\d^3 r_1\d^3 r_2
\notag\\   
   =& \sum_{i=1}^{\norb} \braket{\lmo{i}\lmo{i}|\frac{1}{\norm{\opvec r_1-\opvec r_2}}|\lmo{i}\lmo{i}}
   \label{eq:LocFnEr1}.
\end{align}
For completeness, we also mention the von Niessen (vN) functional\cite{vonNiessen:OrbitalChargeOverlapLocalization1972}
\begin{align}
   L_{\vN}\Bigl[\IndexedSetI{\lmo{i}}{i}\Bigr] :=& \sum_{i=1}^{\norb} \iint \denslmo{i}(\vec r_1)\delta\mleft(\vec r_1-\vec r_2\mright)\denslmo{i}(\vec r_2)\,\d^3 r_1\d^3 r_2
   \label{eq:LocFnVn1}.
\end{align}
vN is harder to motivate than FB or FB, and almost never seen in practice; nevertheless, vN shares most of the formal and practical properties with ER and FB.

In particular, in most molecules, ER, FB, and vN yield visually indistinguishable localization results.
Note that all three functionals are formulated in terms of expectation values of operators which, at least in principle, are physically observable.

In contrast, PM\cite{pipek:PMlocalization} proposed to define the orbital set $\IndexedSetI{\lmo{i}}{i}$'s compound locality in terms of the distribution of gross Mulliken partial charge $Q_A^{i}{}'$ of the orbitals $\lmo{i}$ across the atoms $A$ of the molecule:
\begin{align}
   L_{\PM}\Bigl[\IndexedSetI{\lmo{i}}{i}\Bigr] &:= \sum_{i=1}^{\norb}\sum_{A=1}^{\Nnuc} \bigl(Q_A^{i}{}'\bigr)^2
   \label{eq:LocFnPm}
\end{align}
This criterion is ingenious; it is entirely different from the FB and ER criteria, and not only because it cannot be formulated in terms of an orbital's charge density.
By essentially defining that an orbital is more local if its charge is spread over a smaller number of atoms---regardless of where in space those atoms are located---it introduces an entirely different notion of what locality means.
(and this measure is arguably more relevant to chemical thinking\cite{knizia:iao,knizia:CurlyArrows} than the locality measures of FB and ER).
An excellent comparative technical discussion of all three classical localization methods
can be found in the original PM paper.\cite{pipek:PMlocalization}

Unfortunately, the original PM functional \Eqref{eq:LocFnPm}, by its reference to Mulliken charges, contains a fundamentally unphysical dependence on the used computational basis set. 
This can cause effects ranging from subtle artifacts
to entirely nonsensical results.\cite{knizia:iao,lehtola:GeneralizedPM,knizia:CurlyArrows}

Therefore generalized PM functionals\cite{knizia:iao,lehtola:GeneralizedPM,Senjean:RelativisticIbo2021}
\begin{align}
   L_{\gPM}\Bigl[\IndexedSetI{\lmo{i}}{i}\Bigr] &:= \sum_{i=1}^{\norb}\sum_{F=1}^{N_\mathrm{frag}} h\mleft(\braket{\lmo{i}|\hat n_F|\lmo{i}}\mright),
   \label{eq:LocFnGenPm}
\end{align}
have been proposed, wherein
$\hat n_F$ is a physically well-defined 
\footnote{
Even though the charge operators are now physically well-defined, they still involve \emph{choices} regarding how electronic charge is partitioned among the \emph{interacting} atoms of a molecule (subsystems of an interacting quantum system are not uniquely defined). For example, even when deriving charges from experimentally observable X-ray electron densities of a molecule, one still needs to partition the total electron density into individual atomic contributions, which are not themselves directly observable. How this partitioning is performed is ultimately a matter of choice.}
electron population count operator for fragment $F$
and
$h(n)$ is a driving function (in general, a convex scalar function at least twice-continuously differentiable on $[0,1]$;\cite{Senjean:RelativisticIbo2021} here $h(n)=n^p$ with $p>1$).

As mentioned above, among the classical localization methods, only the PM criterion \Eqref{eq:LocFnPm} yields $\sigma$/$\pi$-separated orbitals. This raises the question of why this separation occurs.
Pipek and Mezey themselves explained this behavior using symmetry arguments based on the Mulliken charge-density operators used in their localization functional.
However, this explanation is not sufficient to account for the more general behavior of PM localization.
 Both the original\cite{pipek:PMlocalization} and generalized\cite{knizia:iao,lehtola:GeneralizedPM} PM methods show $\sigma$/$\pi$-separation even in molecules,  which do not have an exact reflection plane, and where the original symmetry argument does not apply.
Since the PM's\cite{pipek:PMlocalization}  argument relies on a strict classification of orbitals according to their reflection symmetry, the presence of only local or approximate reflection symmetry is \emph{insufficient} to explain the observed $\sigma$/$\pi$-separation in general systems with $\pi$-bonds.

One may hypothesize/conjecture that PM's definition of locality via abstract Hilbert-space charge operators, instead of real-space physical interactions (as in ER and FB), lies at the heart of this difference. 
However, a closer look reveals \emph{another} fundamental difference between the generalized PM (\Eqref{eq:LocFnGenPm}) and the FB, ER, and vN functionals (\Eqsref{eq:LocFnBoys1,eq:LocFnEr1,eq:LocFnVn1}): 
the latter are formulated in terms of \emph{exclusively electronic} properties, while the partial charge operators in the generalized PM functional \Eqref{eq:LocFnGenPm} describe a \emph{relation between electrons and nuclei}.

Based on this insight, we formulated two new localization functionals.
Like PM, these reference the nuclei as an anchor of the localization; but unlike PM, they are constructed in terms of undisputably well-defined physical observables.

First, in the \ndlFullName{} (\textndl{}), we characterize locality by the squared distance of the orbital electron densities (\Eqref{eq:OrbitalElectronDensity}) from the nuclei:
\begin{align}
   L_\mathrm{\ndl}\Bigl[\IndexedSetI{\lmo{i}}{i}\Bigr] 
   :=& -\sum_{A=1}^{\Nnuc}\sum_{i=1}^{\norb} h\mleft(\int \denslmo{i}(\vec r){\norm{\vec r-\vec R_A}^2}\,\d^3 r\mright)
\notag\\   
    =& -\sum_{A=1}^{\Nnuc}\sum_{i=1}^{\norb} h\mleft(\braket{\lmo{i}|{\norm{\opvec r-\vec R_A}}^2|\lmo{i}}\mright).
   \label{eq:LocFnNdl}
\end{align}
As in \Eqref{eq:LocFnGenPm}, $h(x)$ is a convex driving function; here we use $h(x)=x^2$
(initial tests suggest that, just like in gPM,\cite{knizia:iao,lehtola:GeneralizedPM} different choices of $h(x)$ usually yield equivalent localizations).
\Eqref{eq:LocFnNpl} could be interpreted as a nuclear-electronic analog of the FB localization (\Eqref{eq:LocFnBoys1}).
Second, in the \nplFullName{} (\textnpl{}), we characterize locality by the strength of the interaction potential between the the orbital electron densities (\Eqref{eq:OrbitalElectronDensity}) and the nuclei:
\begin{align}
   L_\mathrm{\npl}\Bigl[\IndexedSetI{\lmo{i}}{i}\Bigr] 
   :=& \sum_{A=1}^{\Nnuc}\sum_{i=1}^{\norb} h\mleft(\int \frac{\denslmo{i}(\vec r)\mathinner{Z_A}}{\norm{\vec r-\vec R_A}}\,\d^3 r\mright)
\notag\\   
    =& \sum_{A=1}^{\Nnuc}\sum_{i=1}^{\norb} h\mleft(\braket{\lmo{i}|\frac{Z_A}{\norm{\opvec r-\vec R_A}}|\lmo{i}}\mright).
   \label{eq:LocFnNpl}
\end{align}

\Eqref{eq:LocFnNpl} could be interpreted as a nuclear-electronic analog of the ER localization \Eqref{eq:LocFnBoys1},
as it characterizes locality by the Coulomb interaction potential of the orbital electron densities with the nuclei (\Eqref{eq:LocFnNpl}) instead of with themselves (\Eqref{eq:LocFnBoys1}).

In order to establish the properties of the here-proposed localization functionals in \Eqsref{eq:LocFnNdl,eq:LocFnNpl}, and to perform a fair comparison to the IBO method as well as the classical FB and ER localizations, we implemented solvers for all of those localization methods into a development versions of \progname{IboView} and PySCF.

The implemented solvers are based on a simple iterative algorithm already used in Ref.~\onlinecite{knizia:iao}, but only recently presented with full derivations and technical details (Ref.~\onlinecite{Senjean:RelativisticIbo2021} appendix B).
The algorithm is an improved and generalized version of PMs',\cite{pipek:PMlocalization}
and allows maximizing, in the sense of \Eqref{eq:UnitaryArgmax}, a general functional
\begin{align}
   L_\mathrm{gen}\Bigl[\IndexedSetI{\lmo{i}}{i}\Bigr] &:= \sum_{i=1}^{\norb}\sum_{\ilop=1}^{\Nlop} h\mleft(\braket{\lmo{i}|\lop{\ilop}|\lmo{i}}\mright),
   \label{eq:LocFnGen}
\end{align}
where $\IndexedSetE{\lop{\ilop}}{\ilop}{1}{\Nlop}$ is any set of Hermitian $\Nlop$ operators, and $h$ is the driving function (which is necessary because the direct integral expression is unitarily invariant). The algorithm requires that the driving function $h(x)$ is well-defined and at least twice-continuously differentiable on an interval $D\supset\bigcup_{\ilop}\sigma\mleft(\lop{\ilop}\mright)$ covering the spectra of the $\lop{\ilop}$ operators (i.e., covering all possible $\braket{\phi|\lop{\ilop}|\phi}$).
In order to lead to localization, $h(x)$ should, furthermore, be \emph{strictly convex} on this interval (i.e., have a positive second derivative).
This property is discussed in \appxref{sec:ConvexityCriterion}.

While this fact might not be obvious at first glance, this form allows implementing \emph{all} of the mentioned localization functionals: it covers the PM, FB, \textndl{} and \textnpl{} functionals exactly, and the ER and vN functionals in a robust density-fitting (DF) approximation (see \appxref{sec:implementation}).

\section{Computational details}

For all systems, the reference SCF wavefunction was obtained from either restricted or unrestricted  (in the case of the $[\mathrm{Fe}_2\mathrm{S}_2(\mathrm{SCH}_3)_4]^{3-}$ cluster) density functional theory Kohn-Sham calculations using the PBE \cite {perdew:pbe}  functional and the def2-TZVP \cite{Weigend:def2SVP_def2TZVPP} basis set. Then so obtained molecular orbitals (MOs) were localized with various localization approaches.
We did not use any point group symmetry. All calculations here were performed with the PySCF software package \cite{sun2018pyscf,sun2020recent} in combination with our developmental Python-based code. The orbitals were visualized using the IboView program.  

\section{Results and Discussions}


\subsection{General variational principle behind $\sigma/\pi$ separation }
\label{sec:GenPrinciple}

Localization methods based on real-space electronic observables (FB and ER) optimize geometric compactness rather than chemical locality. Therefore, they often favor ``banana''-shaped orbitals that minimize orbital spread or maximize self-repulsion energy, although these orbitals do not necessarily correspond to chemically meaningful $\sigma$ or $\pi$ bond-like localized orbitals.

Atomic population(charge)-based localization methods (PM, IBO) achieve $\sigma/\pi$-separation by using orbital atomic populations as local descriptors, i.e. using a chemical atom-based notion of locality. The original PM  explanation based on molecular reflection symmetry successfully rationalizes $\sigma/\pi$-separation in molecules with reflection planes, but does not explain the broader practical observation that PM-type and IBO-type localizations preserve $\sigma/\pi$-separation in many systems where exact point-group symmetry is absent or only approximate.\cite{lehtola:GeneralizedPM}

Here, we provide a more general mechanism for $\sigma/\pi$-separation that not only explains why PM localization works, but also shows that atomic populations are not the only local descriptors capable of producing chemically meaningful localization. Our reasoning is based on the convexity of the driving function of the localization functional (see  \appxref{sec:ConvexityCriterion} for mathematical details). In the  context or orbital localization, convexity means that the localization functional energetically favors unequal distributions of local orbital contributions over averaged ones. As a consequence, mixing orbitals with different local character becomes variationally unfavorable, preserving $\sigma/\pi$-separation.

This interpretation also explains the behavior of the classical localization schemes. Since FB and ER functionals are not constructed from convex nonlinear functions of chemically resolved local orbital contributions, they do not variationally suppress $\sigma/\pi$-mixing when hybridized orbitals become geometrically more compact.

Since $\sigma$-orbitals are almost universally strongly localized (in the charge-spread sense) than $\pi$-orbitals, generalized PM-type methods, by their very construction,  disfavor $\sigma/\pi$ mixing. In generalized PM methods, the local orbital contributions correspond to orbital atomic populations, whereas in the NPL and NDL methods presented in this work they are defined through expectation values of electron–nucleus physical operators. The mechanism responsible for $\sigma/\pi$-preservation is therefore does not depend on atomic populations themselves, but on the convex structure of the localization functional.

In general,   $\sigma/\pi$-separation is not determined by whether a localization functional is geometric or observable-based, but rather by whether it uses a convex nonlinear weighting of local orbital contributions that favors distinct localized orbitals over globally averaged/hybridized combinations. $\sigma/\pi$-preserving localization functionals can be constructed from any local orbital quantity that distinguishes different orbital manifolds  through a convex nonlinear transformation. This provides a general route for constructing conceptually different yet chemically meaningful  $\sigma/\pi$-separating localization functionals from other physical observables.

\subsection{Organic molecules with conjugated $\pi$-bonds}
\label{sec:pi_org}

\begin{figure}[h!]
  \centering
  \includegraphics[width=0.9\columnwidth]{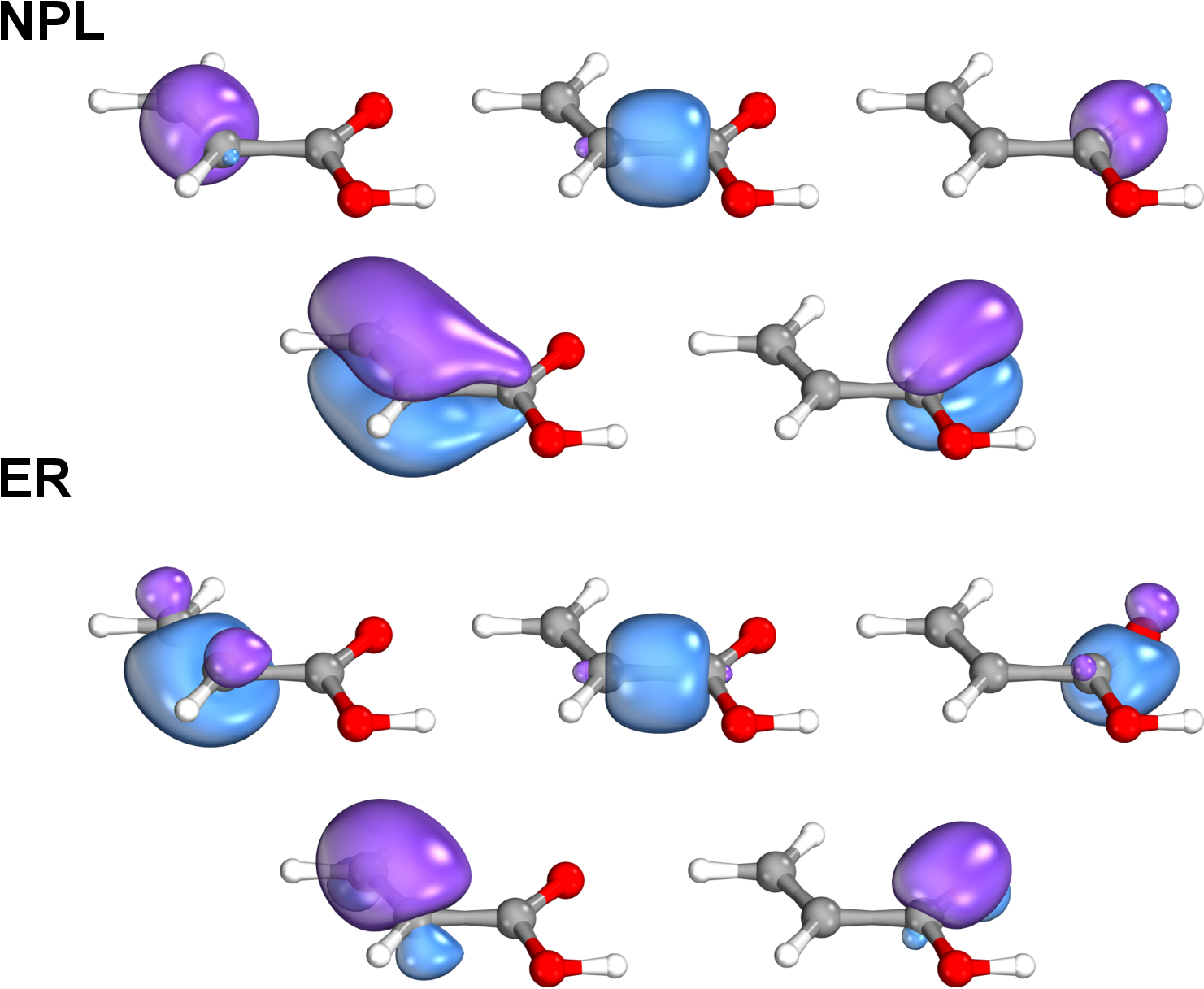}
  \caption{Localized molecular orbitals obtained for acrylic acid with the NPL and ER methods. The orbitals are vizualized with the isosurface threshold corresponding to an 80$\%$ density cutoff}
  \label{fig:acryl}
\end{figure}

We begin by considering simple organic molecules containing both $\sigma$- and $\pi$-bonds, such as acrylic acid and pyridine (Here we do not compare inner-shell orbitals).  \figref{fig:acryl} shows the localized MOs of the acrylic acid molecule produced by the ER and NPL methods. 

The ``banana''-shaped orbitals obtained with the ER and FB methods are visually identical, although their orbital energies differ by $\approx$ 0.03-0.04 a.u. The NPL method, along with IBO and NDL approaches yield separate $\sigma$- and $\pi$-orbitals.
They appear to be visually near-indistinguishable in terms of their shapes, however, thorough analysis reveals small but noticeable differences in orbital composition (in terms of atomic orbitals) and energies. The variations in orbital energies are more pronounced/evident in the case of $\sigma$- orbitals (up to $\approx$ 0.15 a.u.) compared to  $\pi$- orbitals ($<$ 0.01 a.u.) shown in \figref{fig:acryl}. 

\begin{figure}[h!]
  \centering
  \includegraphics[width=0.9\columnwidth]{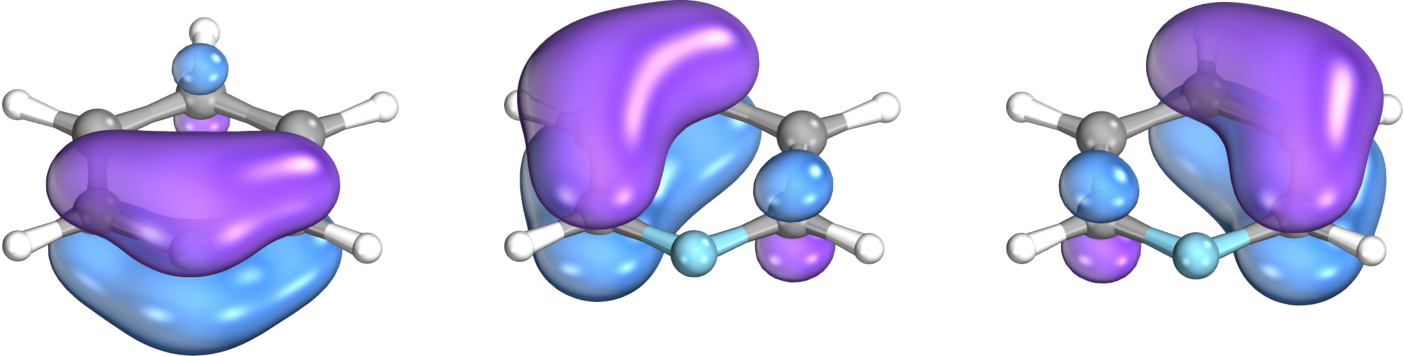}
  \caption{Localized molecular $\pi$-orbitals obtained for pyridine with the NPL method. The orbitals are vizualized with the isosurface threshold corresponding to an 80$\%$ density cutoff. The NDL and IBO methods yield visually near-indistinguishable orbitals. }
  \label{fig:pyridine}
\end{figure}

In the case of pyridine, all of the methods which preserve $\sigma$- and $\pi$-orbitals separation yield visually indistinguishable six $\sigma$-orbitals and three $\pi$-orbitals. 
 \figref{fig:pyridine} displays  $\pi$-orbitals of pyridine obtained with the NPL method.
 The $\pi$-orbital energies obtained using NPL, NDL, and  IBO methods are also very close, with differences being smaller than 0.005 a.u.. Like for acrylic acid, $\sigma$-orbitals have larger differences in energies (up to $\approx$ 0.1 a.u.).

One should exercise caution when using the FB and ER methods, as their localization functionals can be difficult to converge \cite{switkes1970localized,newton1970localized,Kleier:ComparisonERvsBoys1974,rajzmann1987localized},
and even small differences in the resulting values of these functionals can lead to noticeably different orbitals.
This may explain the discrepancies in the ER localization results reported in the literature. While some reseachers claim that the ER can preserve $\sigma$/$\pi$ separation (e.g., Refs.\cite{pipek:PMlocalization,subotnik2007localized, boughton:BPcriterion}), others argue that the ER localization produces ``banana''-shaped orbitals resulting from $\sigma$-$\pi$ mixing. \cite{jensen:IntroToCompChem, marzari2012maximally,Moura2020}. 

The $2 \times 2$ orbital transformation normally employed by the ER method does not necessarily yield the global maximum of the ER functional \cite{rajzmann1987localized}.  When the initial set of orbitals consists of unmixed  $\sigma$ and $\pi$ canonical orbitals, the ER localization 
may preserve $\sigma/\pi$ separation if the functional is not fully maximized. But the global maximum of the ER localization functional instead corresponds to orbitals with mixed $\sigma/\pi$ character in all cases we tested. This can be verified by applying random unitary tranformations to the initial canonical SCF orbitals and subsequently performing ER localization from different initial orbital sets  \cite{Kleier:ComparisonERvsBoys1974}. 

For both pyridine and acrylic acid, the ER localization applied directly to the occupied canonical SCF orbitals, using PySCF’s default convergence criteria, yields well-separated $\sigma$ and $\pi$ orbitals. However, applying the ER localization to orbitals obtained from random unitary rotations of the SCF orbitals produces higher values of the localization functional and the resulting orbitals are localized into “banana” bonds. Specifically, for an 18\textdegree \;  rotation of the initial SCF orbitals (see the details of this unitary rotation in \appxref{sec:URot}), the Coulomb self-repulsion increases by 0.06 a.u. for pyridine and 0.10 a.u. for acrylic acid, demonstrating that the global maximum of the ER functional corresponds to mixed $\sigma/\pi$-orbitals.
For these molecules we have seen the same behavior in every case we tested when ER/FB initially did produce $\sigma/\pi$-separation will default settings of PySCF localization procedure.

We stress that, regardless of this, the FB and ER methods technically yields a physically valid representation of the $N$-electron wave function $\Phi$ (in the sense that the localized orbitals can express $\Phi$ exactly); however, the $\sigma/\pi$ orbital mixing can very negatively impact their applicability for chemical analysis (see SI of Ref.~\onlinecite{knizia:CurlyArrows} for an example).

\subsection{Transition metal complexes} 

The performance of the different localization methods in transition metal complexes significantly depends on the nature of ligands and on the character of their metal–ligand bonding.
In transition  metal complexes also differences between the NDL and NPL methods become evident, which so far have behaved very similar (when applied to organic molecules).

\begin{figure}[h!]
  \centering
  \includegraphics[width=1.0 \columnwidth]{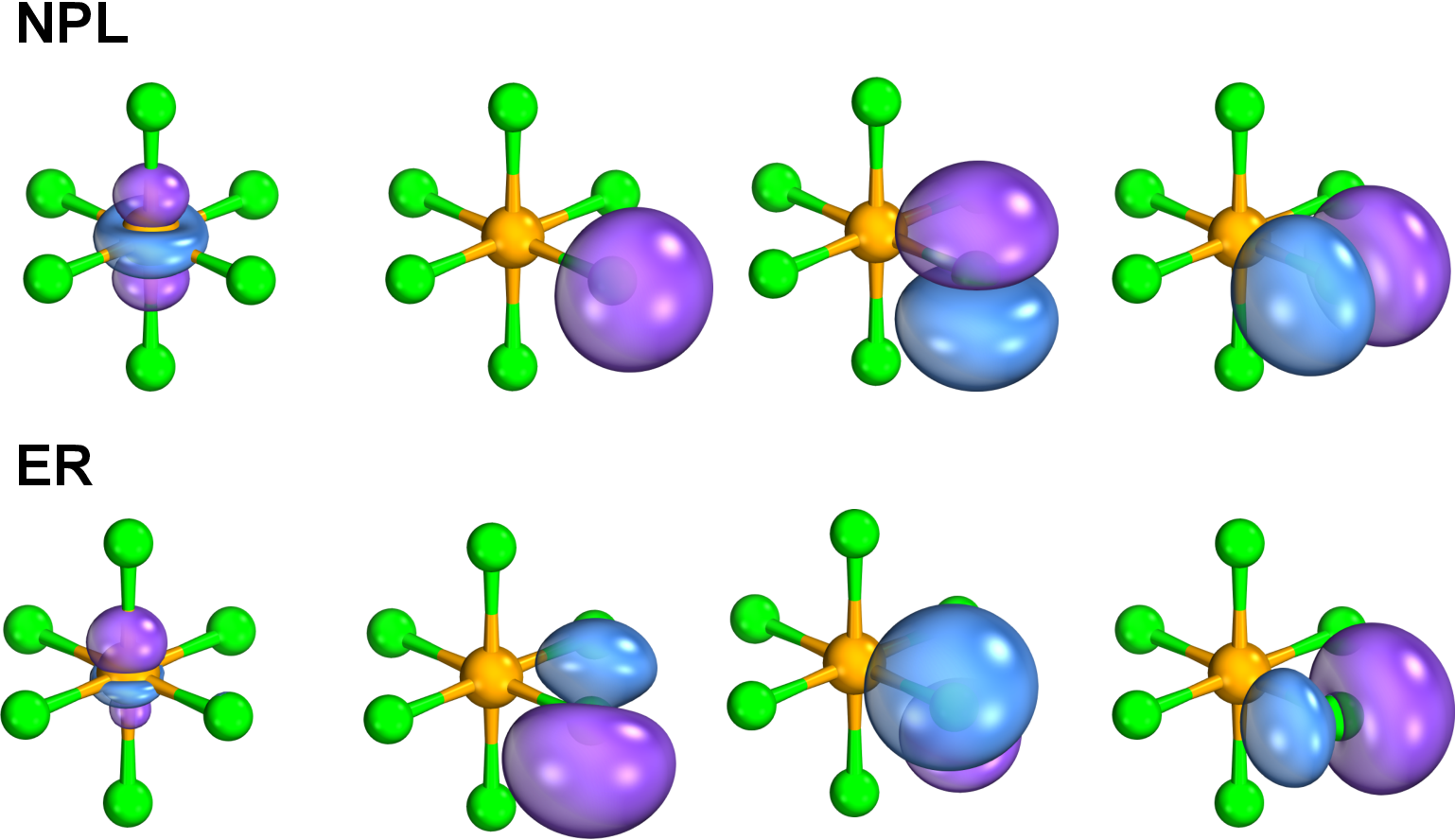}
  \caption{An example of a $3d$-orbital on the Fe center and lone pairs on one of chrloride ligands for \ce{FeCl6^4-} and the hybridized analogs obtained with the ER method. The orbitals are vizualized with the isosurface threshold corresponding to an 92$\%$ density cutoff.}
  \label{fig:fecl6_m4}
\end{figure}

For first row transition metal centers, the localization methods emplyoing real-space orbital compactness criteria (FB and ER) tend to mix 3s, 3p and 3d atomic orbitals into compact but strongly 3s/3p/3d-mixed localized molecular orbitals. 	
As 3s-, 3p-, and 3d-type atomic orbitals are all centered on the same nucleus and have comparable radial extents, their mixing can improve the compactness criterion and is therefore not penalized.
  We illustrate this effect below for a simple  \ce{[FeCl6]^4-} complex (see  \figref{fig:fecl6_m4}), which lacks extended ligands with $\sigma$- and $\pi$ bonding and thus isolates the behavior of the metal center.

The NDL method can also produce mixed (3s/3p/3d) orbitals at transition-metal centers, although the mixing is typically less severe than in the ER and FB methods. Since NDL is formulated in terms of the electron–nucleus distance operator, it is likewise affected by the geometric compactness problem for orbitals with similar radial extents centered on the same atom.

Atomic population(charge)-based localization methods (IBO, PM) , as well as the NPL method, do not demonstrate this type of 3s/3p/3d orbital mixing. 
These methods consistently produce well-defined 3d orbitals; however, in some complexes, may still yield mixed 3s/3p semicore (non-bonding) orbitals.
The 3d orbitals participate anisotropically in metal–ligand bonding and therefore possess significantly different atomic population and  nuclear-potential patterns, making their mixing unfavorable within localization criteria.  In contrast, semicore 3s and 3p orbitals are highly compact, centered on the same metal nucleus, and have very similar atomic population distributions and  nuclear-potential profiles. Rotations within the (3s/3p) semicore orbital manifold only weakly influence the localization functional and are therefore not strongly suppressed.
In practice, this 3s/3p mixing is not problematic, as these chemically inactive semicore orbitals play only a minor role in chemical bonding and reactivity.

One of the main challenges associated with 3s/3p/3d orbital mixing is the reliable identification of the resulting localized orbitals. In some complexes, localized molecular orbitals obtained with localization functionals using real-space electronic observables (ER and FB) retain predominantly 
3p or 3d-character despite partial hybridization, allowing their assignment based on atomic orbital composition or orbital energies (the five predominantly 3d-type orbitals lie higher in energy than the remaining orbitals, which exhibit significant 3s and 3p character). In other cases, however, the degree of mixing is sufficiently strong that no clear orbital character can be assigned. 

\begin{center}
\begin{figure}[h!]
  \includegraphics[width=0.72\columnwidth]{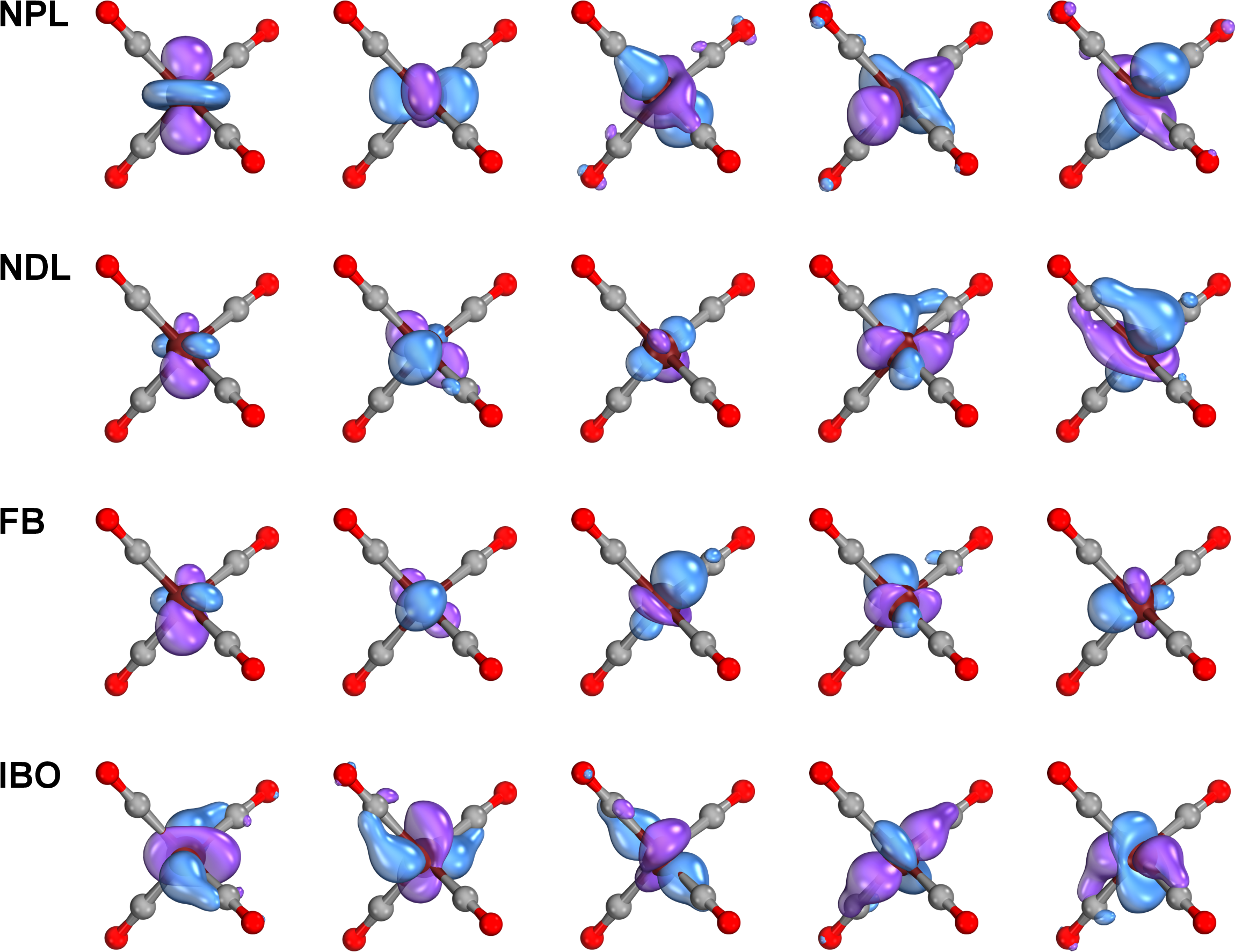}
  \caption{Localized molecular $d$-orbitals obtained for \ce{Ni(CO)4} with different methods after applying an 18\textdegree \;rotation to the initial SCF orbitals. The orbitals are vizualized with the isosurface threshold corresponding to an 80$\%$ density cutoff.}
  \label{fig:NiCO4}
\end{figure}
\end{center}

To illustrate this issue, we consider the \ce{Ni(CO)4} complex. 
The metal-carbonyl complexes are particularly interesting because the metal-carbonyl bond involves both $\sigma$ and $\pi$ interactions; the metal-CO interaction includes the ligand-to-metal forward $\sigma$–donation and the metal-to-ligand $\pi$-backbonding.
 Figure~\ref{fig:NiCO4} visualizes the localized molecular orbitals with predominantly 3d-character obtained for \ce{Ni(CO)4} using the different localization schemes discussed above. 

For this system, the ER method yields three sets of triply degenerate hybridized orbitals, rendering their unambiguous identification nearly impossible; these orbitals are therefore omitted from Fig.~\ref{fig:NiCO4}. The FB method also produces mixed orbitals, although in this case their dominant orbital character can still be reasonably assigned.
The NDL localization yields well-separated $\sigma$ and $\pi$ orbitals on the carbonyl ligands in \ce{Ni(CO)4}, in contrast to the FB and the ER method, while simultaneously producing mixed orbitals on the metal center whose dominant atomic orbital character, however, remains well-defined. 

Atomic population(charge)-based localization methods (PM, IBO) and NPL yield higher quality localized 3d-orbitals for \ce{Ni(CO)4}. In these cases, the identification of localized orbitals is straightforward, both through direct visualization and through analysis of atomic orbital coefficients in the molecular-orbital composition. 

For transition-metal complexes with extended conjugated ligands containing multiple  $\pi$ bonds, such as the bidentate acetylacetonate (acac) ligands considered here, localized orbitals centered on the ligands exhibit the same challenges discussed in Sec.~\ref{sec:pi_org}. Specifically, in the \ce{Co(Acac)3} complex, both the ER and FB localization schemes produce  ``banana'' bonds instead of separate  $\sigma$- and $\pi$-orbitals, whereas the PM and NPL approaches preserve $\sigma/\pi$ separation.

It is interesting that the NDL method maintains $\sigma/\pi$-separation for the C-C-C fragments of the acetylacetonate ligands but yields “banana” bonds for the C–O bonds. In the acetylaceton molecule itself, however, NDL produces separate  $\sigma$- and $\pi$-orbitals on all molecular fragments including the C–O bond.

Therefore, we attribute the NDL results for the \ce{Co(Acac)3} complex to the proximity of the oxygen atoms to the metal center. The interaction with the Co center gives the C--O bond a partial three-center character and polarizes both the  $\sigma$(CO)- and $\pi$(CO) orbitals toward the metal. As a result, the two orbitals become more similar with respect to the electron–nucleus distance operator, reducing the ability of the NDL functional to distinguish between them.

\onecolumngrid
\twocolumngrid

\cleardoublepage

\begin{figure}[h!]
  \centering
  \includegraphics[width=0.95\columnwidth]{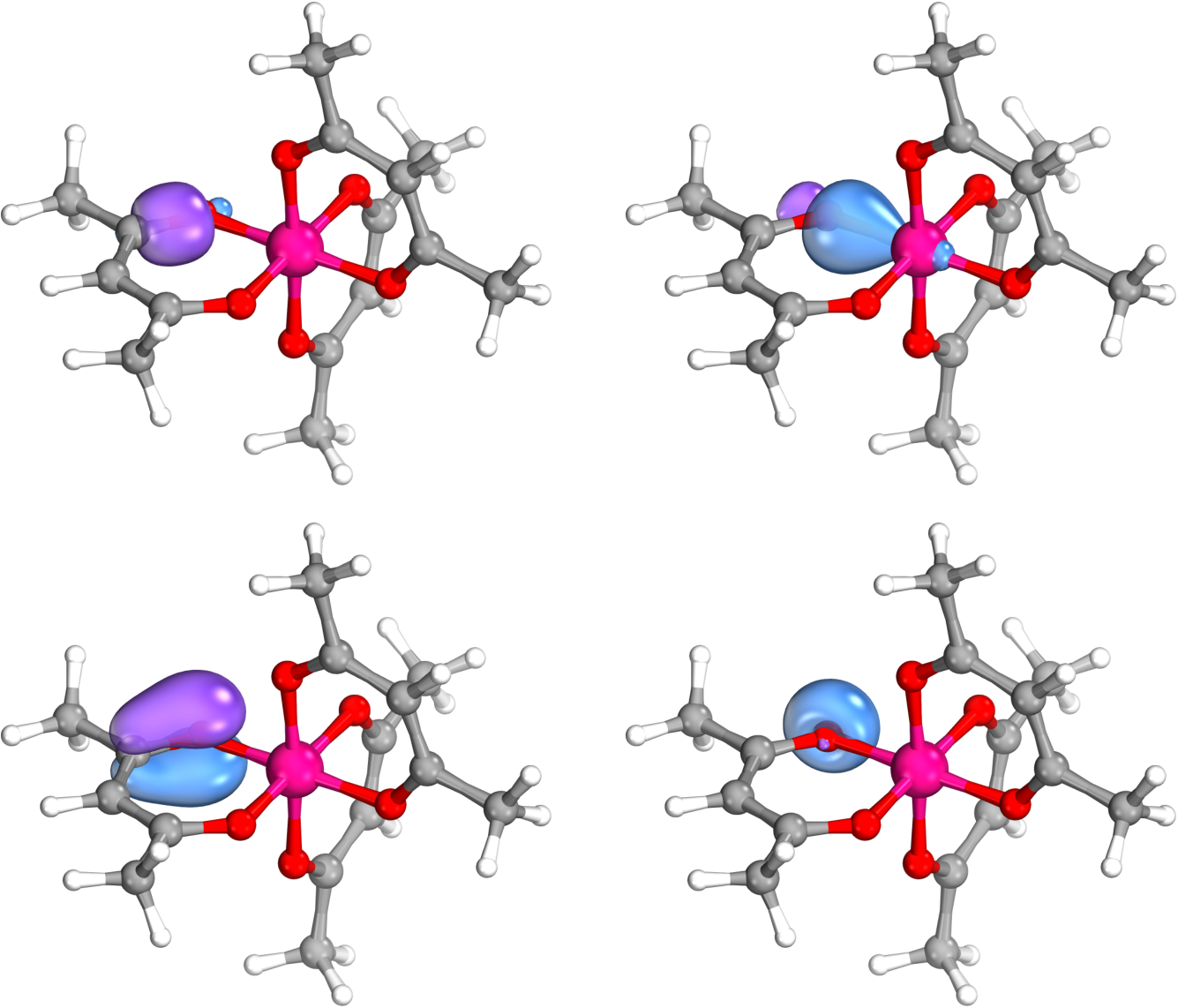}
  \caption{Localized orbitals involving one of O atoms of the acetylacetonate ligand in \ce{Co(Acac)3} obtained with the NPL method. The orbitals are vizualized with the isosurface threshold corresponding to an 80$\%$ density cutoff}
  \label{fig:CoAcac3}
\end{figure}

\figref{fig:CoAcac3} shows the localized orbitals centered on one of the oxygen atoms in an acetylacetonate ligand of the \ce{Co(Acac)3} complex obtained using the NPL method. In planar acetylacetonate ligands, each oxygen atom possesses a single lone pair.

The presence of multiple lone pairs represents another well-known difficulty for localization methods. The ER and FB typically yield hybridized ``rabbit-ear'' orbitals in such cases (see Fig.~\ref{fig:fecl6_m4}).
Atomic population(charge)-based localization methods suffer from a different limitation: in orbital subspaces that are equivalent under the localization functional, such as core orbitals or sets of equivalent lone pairs, the orbital orientations are not uniquely determined.
An arbitrary unitary rotation within these subspaces leads to another set of localized MOs for lone pairs of the same type,
which might be not chemically meaningful.
For example, in the case of \ce{H2O}, instead of two chemically intuitive lone pairs approximately aligned along tetrahedral directions, the localization may yield two rotated lone pairs oriented in an arbitrary plane. In some cases, symmetry-breaking lone-pair orbitals are obtained despite the symmetric molecular geometry. 

Our new nuclear-potential localization (NPL) method overcomes this limitation and eliminate random rotations within degenerate orbital subspaces. The robustness of the method was validated across dozens of transition-metal complexes with diverse geometrical and electronic structures, including metallocenes and metal porphyrins.

Our last example here is the \ce{[Fe2S2(SCH3)4]^3-} complex, that represents a model of the active site in certain iron-sulfur proteins,
such as the ferredoxins in their reduced form.\cite{venkateswara2004synthetic} The ground state of this complex has $S=1/2$ resulting from antiferromagnetic coupling of d-electrons on different iron center. The Fe(III) center should have shorter bond distances with the bridging S atoms and have five $\uparrow$ electrons, while the Fe(II) center should have longer Fe-S distances and six electrons with four unpaired $\downarrow$-electrons among them. \cite{shoji2007theory}.  

\begin{center}
\begin{figure}[h!]
  \includegraphics[width=0.85\linewidth]{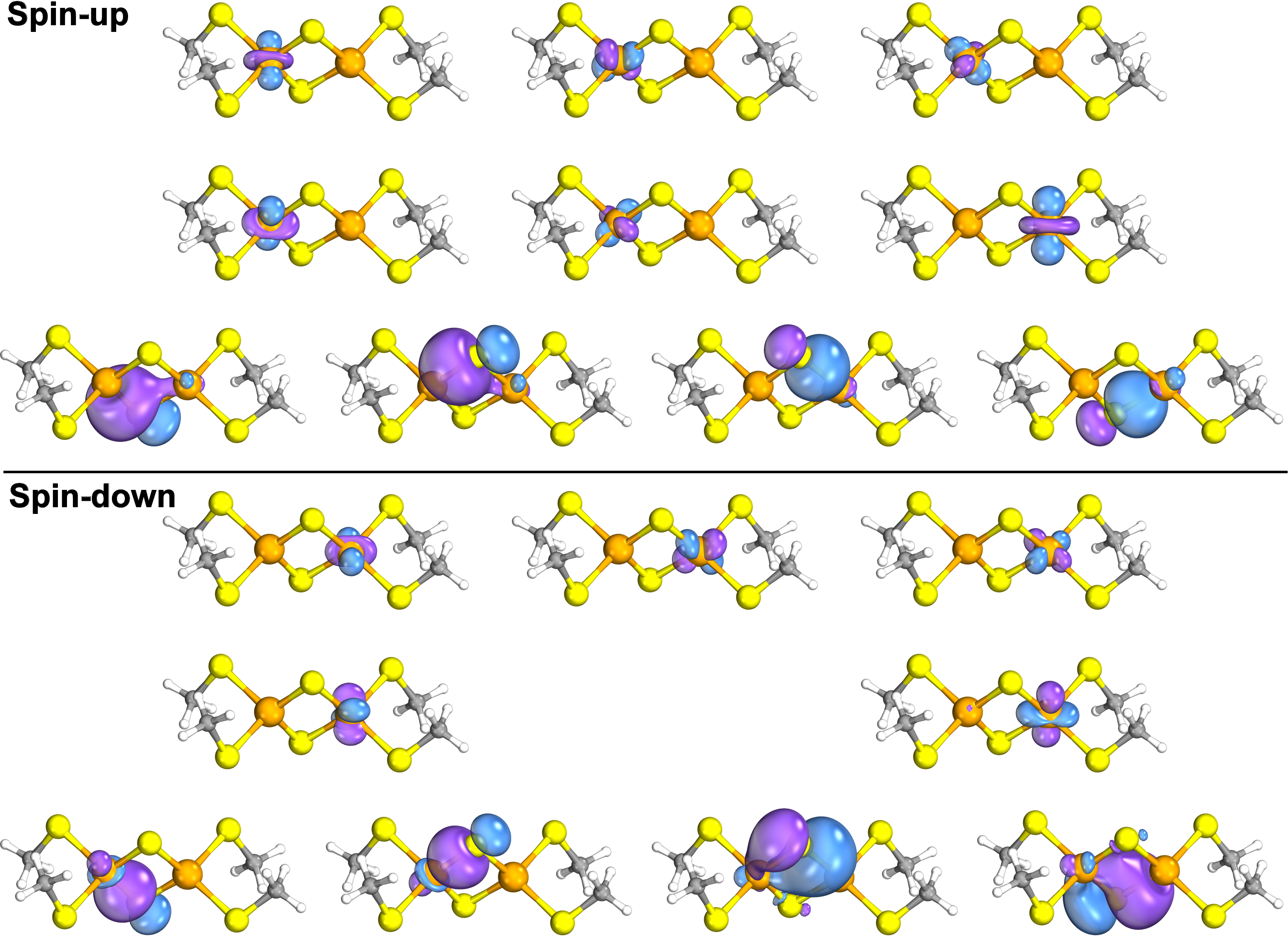}
  \caption{Localized molecular orbitals for the described broken-symmetry UKS state modeling the antiferromagnetic S=1/2 ground state of  \ce{[Fe2S2(SCH3)4]^3-} obtained with the NPL method. First three rows show MOs for $\uparrow$ electrons, and the other rows show MOs for $\downarrow$ electrons. The orbitals are vizualized with an isosurface threshold corresponding to an 80$\%$ density cutoff}.
  \label{fig:Fe2S2}
\end{figure}
\end{center}

To obtain the broken-symmetry reference SCF wavefunction for this complex, we first carried out an unrestricted Kohn-Sham (UKS) calculation on the high spin state with $S = 9/2$, and flipped the spin-density on the iron center that has longer bond distances with the bridging S atoms, and converged the second UKS calculation. Then the so-obtained molecular orbitals were localized.  \figref{fig:Fe2S2} visualizes the Fe--S  bond orbitals and 3d orbitals on the metal centers obtained with the NPL method.

Unlike in many other complexes, the $\uparrow$ and $\downarrow$ spin orbitals here do not form (nearly) identical closed-shell pairs but instead differ noticeably in their spatial forms. However, localization does not break spin symmetry — it only reveals the asymmetry already present in the UKS solution. Although the broken-symmetry solution corresponds overall to an antiferromagnetically coupled low-spin state, the unrestricted formalism allows the $\uparrow$ and $\downarrow$ orbitals to respond to different local spin densities on the two iron centers, resulting in non-identical spatial orbitals even for nominally paired electrons.

\section{Conclusions}\label{sec:summary}

In this work, we report the construction of a new class of orbital localization functionals, and associated numerical algorithms to optimize orbitals with respect to those functionals, which yield $\sigma$-$\pi$-separated orbitals. These functionals are constructed in terms of \textit{physically directly observable} nucleus-electron operators.

In all systems tested in the work,  our new NPL method appears to resolve many of the issues arising with other localization functionals. It not only preserves the separation of $\sigma$- and $\pi$-orbitals , but also lifts degeneracy associated with identical charge distributions. 

The NDL method performs better than other localization approaches based on real-space orbital compactness criteria, such as FB and ER, but it has a system-dependent limitation. In purely organic molecules, the NDL method cleanly separates $\sigma$- and $\pi$-orbitals. However, in transition metal complexes it can mix 3s, 3p and 3d orbitals at the metal centers producing hybridized localized orbitals. In addition,  it may occasionally mix $\sigma$- and $\pi$-orbitals centered on ligand atoms bound to the metal, when both become polarized toward the same metal atom.

We also identified the general variational principle behind $\sigma$/$\pi$-separating bond-orbital localization. 
This principle not only explains the behavior of classical localization methods, but also reveal that generalized PM methods belong to a broader class of localization functionals whose behavior is determined by the convex nonlinear weighting of local orbital descriptors.
The NPL and NDL functionals introduced in this work provide two representative examples of this class, in which the local descriptors are expectation values of atom-centered operators.  Their purpose is therefore not to establish the nuclear potential or radial nuclear distance as optimal descriptors of chemical locality, but rather to demonstrate that chemically meaningful localization can be achieved using physically local observables beyond atomic populations


\onecolumngrid

\twocolumngrid
\bibliography{refs}

\cleardoublepage
\appendix
\def\thesection{\Alph{section}}
\def\thesubsection{\Alph{section}.\arabic{subsection}}
\def\thesubsubsection{\Alph{section}.\arabic{subsection}.\arabic{subsubsection}}
\makeatletter
\renewcommand{\p@subsection}{}
\renewcommand{\p@subsubsection}{}
\makeatother

\newcommand{\appendixsectionbreak}{
   \begin{widetext}
   \noindent\rule{\columnwidth}{2pt}
   \\[-8pt]\noindent\rule{\columnwidth}{1pt}
   \end{widetext}
}

\titleformat{\section}{\raggedright\bfseries\sffamily}{Appendix \thesection:}{0.5em}{}
\renewcommand{\tocname}{List of Appendices}
\section{General Jacobi-rotation based localization algorithm}\label{sec:GeneralTwoByTwoLocalization}

To optimize the listed localization functionals, we employ a simplified generalization (Ref.~\onlinecite{Senjean:RelativisticIbo2021} Appendix B) of PMs' algorithm,\cite{pipek:PMlocalization} which is itself a generalization of ER's algorithm\cite{edmiston:LocalizedAtomicAndMolecularOrbitals} of localization by incremental $(2,2)$-shape Jacobi rotations.
The algorithm divides the optimization of \Eqref{eq:LocFnGen} over all orbitals $\IndexedSetE{\lmo{i}}{i}{1}{\Nloc}$
into a series of two-orbital optimizations of
\begin{align}
   L(\theta)=\sum_{\ilop=1}^{\Nlop}\mleft[\fn{h}{\braket{\lmo{i}|\lop{\ilop}|\lmo{i}}} + \fn{h}{\braket{\lmo{j}|\lop{\ilop}|\lmo{j}}}\mright]\label{eq:TwoOrbitalL}
\end{align}
with respect to the $(2,2)$-shape unitary transformation
\begin{align}
   \lko{i}
   &:= \cos(\theta)\lki{i} + \sin(\theta)\lki{j}\nonumber\\
   \lko{j}
   &:=-\sin(\theta)\lki{i} + \cos(\theta)\lki{j}.
   \label{eq:UTwo}
\end{align}
\Eqref{eq:TwoOrbitalL} is maximized%
   \footnote{If a functional $L$ should be \emph{minimized} instead of maximized, one may either replace \Eqref{eq:MaxLocFnArctan2Arg} by $\theta=\frac{1}{4}\operatorname{arctan2}\mleft(-\,B_{ij}(z),A_{ij}(z)\mright)$ or simply maximize the negated functional $-L$ (as done in this work).}
exactly ($h(x)=x^2$ or $h(x)=x^3$) or approximately (other $h(x)$) by\cite{Senjean:RelativisticIbo2021}
\begin{align}
   \theta = \efrac{1}{4}\operatorname{atan2}\mleft(B_{ij},\;-\,A_{ij}\mright),\label{eq:MaxLocFnArctan2Arg}
\end{align}
where $\operatorname{atan2}(y,x)$ is the two-argument arc tangent function,\footnote{The two-argument arc tangent function $\operatorname{atan}(y,x)$ returns the angle $\theta\in(-\pi,\pi]$ between positive $x$-axis and a ray from $(0,0)$ to $(x,y)$. For $x>0$, it evaluates to $\operatorname{atan2}(y,x)=\operatorname{atan}(y/x)$, and otherwise takes the sign of $x$ and $y$ into account to yield a unique arc angle in the full range of $(-\pi,\pi]$.} 
and $B_{ij}$ and $A_{ij}$ arise from 
the Taylor expansion of 
$L(\theta)$
for $\theta(x)=\efrac{1}{4}\operatorname{atan}\mleft(x\mright)$ around $x\approx0$:
\begin{align}
   L\bigl(\efrac{1}{4}\operatorname{atan}\mleft(x\mright)\bigr) = \frac{1}{2} A_{ij}\,x^2 + B_{ij}\,x
   + C_{ij} + \mathcal{O}(x^3). \label{eq:LxExpansion2ndOrder}
\end{align}
The coefficients, of which only $A_{ij}$ and $B_{ij}$ are needed in \Eqref{eq:MaxLocFnArctan2Arg}, can be evaluated\cite{Senjean:RelativisticIbo2021} for general $h(x)$:
\newcommand{\fnop}[1]{\mathop{{}#1}}  
\newcommand{\fnArg}[2]{\fnop{#1}\mleft(#2\mright)}
\newcommand{\fnarg}[2]{\fnop{#1}(#2)}
\newcommand{\fnargb}[2]{\fnop{#1}\bigl(#2\bigr)}
\newcommand{\fnargB}[2]{\fnop{#1}\Bigl(#2\Bigr)}
\begin{align}
   A_{ij}&=\sum_{\ilop=1}^{\Nlop}\bigg[\efrac{1}{4} \bigl({\LopMel{i}{j}{\ilop}}\bigr){}^2 \left(\fnargb{h''}{\LopMel{i}{i}{\ilop}}+\fnargb{h''}{\LopMel{j}{j}{\ilop}}\right)
\notag\\&\qquad\quad
   -\efrac{1}{8}\bigl (\LopMel{i}{i}{\ilop}-\LopMel{j}{j}{\ilop}\bigr) \left(\fnargb{h'}{\LopMel{i}{i}{\ilop}}-\fnargb{h'}{\LopMel{j}{j}{\ilop}}\right)\bigg]
\notag
\\ B_{ij}&=\sum_{\ilop=1}^{\Nlop}\bigg[\efrac{1}{2} \LopMel{i}{j}{\ilop} \left(\fnargb{h'}{\LopMel{i}{i}{\ilop}}-\fnargb{h'}{\LopMel{j}{j}{\ilop}}\right)\bigg]
\\ C_{ij}&=\sum_{\ilop=1}^{\Nlop}\Bigl[\fnargb{h}{\LopMel{i}{i}{\ilop}}+\fnargb{h}{\LopMel{j}{j}{\ilop}}\Bigr]
\label{eq:UnitaryTwoByTwoABC}
\end{align}

Herein $h'(x)$ and $h''(x)$ denote the first and second derivative of $h(x)$, and the 
operators $\IndexedSetI{\lop{\ilop}}{\ilop}$ of \Eqref{eq:LocFnGen} are represented by their matrix elements
\begin{align}
   \LopMel{i}{j}{\ilop} := \braket{\lmi{i}|\lop{\ilop}|\lmi{j}}.
\end{align}

The pseudocode for the localization algorithm is given in  \figref{fig:UandK}. this is incrementally applied on all unique $(i,j)$ pairs.

\begin{figure}[h!]
Algorithm for Jacobi 2 $\times$ 2 rotations to maximize $L$ from \Eqref{eq:LocFnGen}.
\newcommand{\algocomment}[1]{\color{Gray}\text{\# #1}}
   \rule{\columnwidth}{1pt}
   \begin{align*}
   &\algocomment{Initialize the unitary transformation matrix via the Identity matrix}\\
   &\mat U:= \mat I \\
  &\algocomment{Initialize the $(N_\mathrm{nuc},N_\mathrm{orb},N_\mathrm{orb})$-shaped kernel matrix}\\
  & \mat K  \leftarrow \mat K^{(0)} \\
& \mathbf{for } \; n_\mathrm{iter} \;\mathbf{ in \; range} (N_\mathrm{maxiter}):\\
 & \quad \algocomment{Set the global gradient norm Grad=$||\nabla L||$}\\
  & \quad \mathrm{Grad} = 0\\
& \quad\mathbf{for } \; i \;\mathbf{ in \; range} (N_\mathrm{orb}):\\
& \qquad\mathbf{for } \; j \;\mathbf{ in \; range} (i): \algocomment{$i>j$}\\
& \quad\qquad\mathbf{for } \; F \;\mathbf{ in \; range} (N_\mathrm{nuc}): 
&\quad \qquad\algocomment{Extract the matrix elements and compute the coefficients} \\
 &\qquad\qquad  O_{ii}^F = K_{A,ii}  \\
 &\qquad\qquad  O_{jj}^F = K_{A,jj} \\
 &\qquad\qquad   O_{ij}^F =K_{A,ij} \\
&\qquad\qquad A_{ij} = \sum_A \left[ 4(O_{ij}^A)^2 - (O_{ii}^A - O_{jj}^A)^2 \right] \\
& \qquad\qquad B_{ij} = \sum_A 4\,O_{ij}^A (O_{ii}^A - O_{jj}^A)\\
&\quad\qquad\algocomment{Determine the rotation angle and apply the Jacobi rotation}\\
&\qquad\qquad \theta_{ij} = \frac{1}{4}\,\arctan2\!\left(B_{ij} h_\mathrm{sign},\,-A_{ij} h_\mathrm{sign}\right)\\
 &\quad\qquad\algocomment{$h_\mathrm{sign}=1$ selects maximization  of L} \\
&\quad\qquad\algocomment{$h_\mathrm{sign}=-1$ selects minimization of L} \\
&
\qquad\qquad  \left [\begin{matrix}
U_{ik} \\ 
U_{jk}  
\end{matrix}\right ]  \leftarrow \left [\begin{matrix}
\cos (\theta_{ij} )& -\sin ( \theta_{ij} ) \\ 
\sin ( \theta_{ij} )& \cos (\theta_{ij} )
\end{matrix}\right ]
\left [\begin{matrix}
U_{ik} \\ 
U_{jk}  
\end{matrix}\right ]=
\left [\begin{matrix}
cs& -ss\\ 
ss& cs
\end{matrix}\right ]
\left [\begin{matrix}
U_{ik} \\ 
U_{jk}  
\end{matrix}\right ]\\
&\quad\qquad\algocomment{ update the corresponding kernel sub-block}\\
&\qquad\qquad\left [\begin{matrix}
K_{A,ii}&K_{A,ij} \\ 
K_{A,ii}& K_{A,jj}
\end{matrix}\right ] \leftarrow
\left [\begin{matrix}
cs& ss\\ 
-ss& cs
\end{matrix}\right ]
\left [\begin{matrix}
K_{A,ii}&K_{A,ij} \\ 
K_{A,ii}& K_{A,jj}
\end{matrix}\right ] 
\left [\begin{matrix}
cs& -ss\\ 
ss& cs
\end{matrix}\right ]\\
&\quad\qquad\algocomment{ update the gradient}\\
&\qquad \mathrm{Grad} \leftarrow \mathrm{Grad} +(4B_{ij}\theta_{ij} )^2\\
&\quad\qquad\algocomment{$B_{ij}$ is the actual gradient. }\\
&\quad\qquad\algocomment{$A_{ij}$ is effectively the second derivative at $\theta = 0$.}\\
&\quad\mathrm{Grad} \leftarrow \sqrt{\mathrm{Grad}}\\
&\quad\mathbf{if} \; \mathrm{ Grad} <\mathrm{ThresholdGrad} \\
&\qquad \mathbf{break}    \\
&\mathbf{return} \; \mat U
   \end{align*}
   \caption{Numerical algorithm for optimizing the unitary transformation matrix $\mat U$ and the kernel $\mat K$ with inline rotations.
 Note that these rotations are incrementally applied to all unique $(i,j)$ pairs. In the current code, we keep a local copy of $\mat K$ and update it whenever the orbitals are rotated in the 2x2 sweeps}
   \label{fig:UandK}
   \rule{\columnwidth}{1pt}
\end{figure}
\cleardoublepage

\section{Algorithm details for the \textnpl{}, \textndl{}, IBO, FB, and ER localizations} \label{sec:implementation}

We here describe concrete technical details relevant to the optimization of the various localization functionals described in the text.
While, as we will show, the algorithm of \secref{sec:GeneralTwoByTwoLocalization} can be applied as-is to all the titular localizations (at least in some variants), in our previous work\cite{knizia:iao,Senjean:RelativisticIbo2021} we were concerned with Intrinsic Bond Orbitals (IBOs) only and did not recognize or discuss its full generality.

\subpoint{\textndl{} and \textnpl{} localizations}
The localization functionals \Eqref{eq:LocFnNdl} (for \textndl) and \Eqref{eq:LocFnNpl} (\textnpl) are already written in the form of 
\Eqref{eq:LocFnGen}.
The algorithm of \secref{sec:GeneralTwoByTwoLocalization} can be applied as-is if we identify $\IndexedSetI{\lop{\ilop}}{\ilop}$ with the $\Nlop:=\Nnuc$ individual nuclear interaction operators.
The $(\Nnuc,\Nloc,\Nloc)$-shape tensors of localization operator $\lop{\ilop}$ matrix elements are
given by
\newcommand{\FragSet}[1]{\mathcal{F}_{#1}}
\begin{align}
   \LopMel{i}{j}{\ilop} &:= \sum_{A\in \FragSet{\ilop}}\braket{\lmi{i}|\norm{\opvec r - \vec R_A}^2|\lmi{j}}
   &&\text{(for \textndl)}
\label{eq:NdlLopMel}
\\
   \LopMel{i}{j}{\ilop} &:= \sum_{A\in \FragSet{\ilop}}\braket{\lmi{i}|\frac{Z_A}{\norm{\opvec r - \vec R_A}}|\lmi{j}}
   &&\text{(for \textnpl)}
\label{eq:NplLopMel}
\end{align}
The molecular integrals occurring in these matrix elements can be computed from the $xx$, $yy$, and $zz$-components of the Cartesian quadruple/second moment integrals 
$(\chi_a|\hat r_{\kappa}\hat r_{\lambda}|\chi_b)$
(for \Eqref{eq:NdlLopMel}) or nuclear attraction integrals (for \Eqref{eq:NplLopMel}); both of those are susceptible to standard Gaussian integration methods (e.g., chapters 9.3.2 and 9.10.1 in Ref.~\onlinecite{helgaker:purplebook}) and readily available in most quantum chemistry programs.
Alternatively, they can be efficiently computed as degenerate two-electron three-center integrals\cite{ahlrichs:3centerintegrals,ahlrichs:osrr} $(\chi_a\chi_b|K(r_{12})|\chi_c)$ with $K(r_{12})=r_{12}^{-1}$ or $K(r_{12})=r_{12}^2$ in the point-charge limit of $s$-type Gaussian functions $\chi_c(\vec r)\propto \Exp{-\gamma\smash{(\vec r-\vec R_A)^2}}$ placed at the nuclei $\vec R_A$ (see Eqs. (A8) and (F20) in the supporting information of Ref.~\onlinecite{peels:ir2c}).
We implemented this latter route.

\subpoint{Boys localization}
The Boys functional $L_{\FB}$ in \Eqref{eq:LocFnBoys1} can be reformulated into several equivalent forms (see Ref.~\onlinecite{pipek:PMlocalization} near eqs.~(3a)--(7) for a discussion); one of these\cite{pipek:PMlocalization} is
\begin{align}
   L_{\FB}\Bigl[\IndexedSetI{\lmo{i}}{i}\Bigr] &=
    2\sum_{i=1}^{\norb} \Bigl(\sum_{\kappa\in\{\mathrm{x},\mathrm{y},\mathrm{z}\}}\,\abs{\braket{\lmo{i}|\hat r_\kappa|\lmo{i}}}^2 - \braket{\lmo{i}|\norm{\vec r}^2|\lmo{i}}\Bigr),
   \label{eq:LocFnBoys2}
\end{align}
and this is the form we implemented (see also the explanation in \appxref{sec:diffBoys}).
Note that the $\norm{\vec r}^2$ terms of \Eqref{eq:LocFnBoys2} can be simply discarded, because $\sum_i \braket{\lmo{i}|\hat r^2|\lmo{i}}$ is invariant to unitary transformations between the $\{\lmo{i}\}$.

The remaining terms then fit into the framework of \Eqref{eq:LocFnGen} if we use as driving function $h(x)=x^2$ and as $\IndexedSetI{\lop{\ilop}}{\ilop}$ the $\Nlop=3$ Cartesian first-moment/dipole operators $\hat r_\kappa$ ($\kappa\in\{x,y,z\}$).
The $(3,\Nloc,\Nloc)$-shape tensor of localization operator $\lop{\ilop}$ matrix elements is then given by
\begin{align}
   \LopMel{i}{j}{\kappa} &:= \sum_{A\in \FragSet{\ilop}}\braket{\lmi{i}|\hat r_\kappa|\lmi{j}}
   &&\text{(for Boys)}
\label{eq:NdlLopMel}
\end{align}

\subpoint{Edminston-Ruedenberg localizations}

Compared to the other localization methods, ER localizations are \emph{very} expensive to compute. 
The original ER method scales as $N^5$ with the number of basis functions (or atomic orbitals) for a molecule, while PM and FB scale as $N^3$. 
For large molecule, the time of the ER localization significantly exceeds the time of the preceding SCF calculation.
This renders it impractical for chemical analysis, even though it might seem fundamentally better as it is based on the energy condition rather than an ill-defined concept of ``atomic charges'' (PM) or the distance criterion (FB). 
The density fitting(DF)-based implementation of the ER localization can accelerate its performance. Thus, we implement the ER method in the DF approximation.

\section{Convexity of the driving function $h(x)$}\label{sec:ConvexityCriterion}
We stated that, in order to lead to localization if the $L$ of \Eqref{eq:LocFnGen} is maximized, the driving function $h(x)$ should be convex.
A function $h(x)$ is called strictly convex on the interval $D$ if for all $a,b\in D$ and all $\alpha\in(0,1)$
\begin{align}
h\bigl(\alpha\,a+(1-\alpha)\,b\bigr)< \alpha\,h(a) + (1-\alpha)\,h(b).
\label{eq:ConvexFunctionCondition}
\end{align}
If $h(x)\in C^2(D,\mathbb{R})$ (i.e., $h'(x)$ and $h''(x)$ exist and are continuous at all $x\in D$), then $\forall x\in D:\,h''(x)>0$ is a sufficient condition.

Why the convexity criterion?

Consider a two-orbital optimization, in which, leaving all other orbitals unchanged, $(\lmi{i},\lmi{j})$ are rotated into $(\lmo{i},\lmo{j})$ (\Eqref{eq:UTwo}), and define
\begin{align}
   a &:=\braket{\lmi{i}|\lop{\ilop}|\lmi{i}} & b &:=\braket{\lmi{j}|\lop{\ilop}|\lmi{j}} 
\notag\\ 
   a'&:=\braket{\lmo{i}|\lop{\ilop}|\lmo{i}} & b'&:=\braket{\lmo{j}|\lop{\ilop}|\lmo{j}}.
\end{align}
So by the transformation, the contribution to the localization functional $L$ (\Eqref{eq:TwoOrbitalL}) changes from $h(a)+h(b)$ to $h(a')+h(b')$.
However, as shown below, the convexity of $h$ implies
\begin{align}
h(a) + h(b) &< h(a') + h(b') \;\;\;\text{iff}\;\;\; \abs{a-b}<\abs{\smash{a'-b'}}.
\label{eq:ConvexHSumAB}
\end{align}
This means that, at the level of each individual $\lop{\ilop}$ and each pair of orbitals,
a convex $h(x)$ will drive the optimization towards an increasing absolute difference between $\braket{\lmo{i}|\lop{\ilop}|\lmo{i}}$ and $\braket{\lmo{j}|\lop{\ilop}|\lmo{j}}$ (meaning one gets larger and the other one smaller; note that if the $\lop{\ilop}$ are bounded from below/above, the contributions cannot get arbitrarily small/large).
Ultimately, the overall process is therefore typically driven towards each orbital $\lmo{i}$ having large absolute matrix elements $\braket{\lmo{i}|\lop{\ilop}|\lmo{i}}$ for a small number of operators $\lop{\ilop}$ and very small absolute matrix elements for all others.
Due to the nature of this process, in the majority of molecules with well-defined dominant Lewis structures of two-center bonds, most concrete choices of $h(x)$ will yield near-indistinguishable localizations.
However, qualitative differences may arise in molecules which feature bonding with significant multi-center character.

Note that for some combinations of a studied system and other forms of $h(x)$ one can observe a pathological and erratic behavior. However, this is beyond the scope of this paper and will be discussed in detail in our upcoming manuscript.

\Eqref{eq:ConvexHSumAB} can be verified as follows:
As $\sum_{i}\braket{\lmo{i}|\lop{\ilop}|\lmo{i}}$ is unitarily invariant, we have
\begin{align}
   a' + b' &= a + b.           \label{eq:ConvexSumAB}
\end{align}
By \Eqref{eq:ConvexHSumAB}, $[a,b]$ and $[a',b']$ have the same center
$a/2+b/2$%
; so if $\abs{a-b}<\abs{\smash{a'-b'}}$, 
then $a\in[a',b']$, 
and therefore an $\alpha\in(0,1)$ exists with
\begin{align}
a=\alpha\,a' + (1-\alpha)\,b'.
\end{align}
Since $a+b=a'+b'$, the same $\alpha\in(0,1)$ also yields
\begin{align}
   b=(1-\alpha)\,a' + \alpha\,b'.
\end{align}

Let $\beta := 1- \alpha$ and note that $\beta\in(0,1)$ because $\alpha\in(0,1)$.
The convexity condition \Eqref{eq:ConvexFunctionCondition} then yields
\begin{align}
   h(a) &= h\bigl(\alpha\,a' + (1-\alpha)\,b'\bigr)
   < \alpha\,h(a') + (1-\alpha)\,h(b')
\notag\\
   h(b) &= h\bigl(\beta\,a' + (1-\beta)\,b'\bigr)
   < \beta\,h(a') + (1-\beta)\,h(b').
\end{align}

To obtain \Eqref{eq:ConvexHSumAB}, we just add both inequalities:
\begin{align}
   h(a) + h(b) &< (\alpha+\beta)\,h(a') + (1-\alpha+1-\beta)\,h(b')
\notag\\&
               = h(a') + h(b'). \label{eq:h_ineq}
\end{align}


The \Eqref{eq:h_ineq} provides a mathematical explanation for the separation of $\sigma$- and $\pi$-orbitals. A pair consisting of one very localized ($\sigma$) orbital and one less localized ($\pi$) orbital yields a larger value of the localization functional $L$ than two equally localized (``banana''-type) orbitals. 
This is also the fundamental reason why PM localization works. 
Generalized PM-type methods by construction favor orbitals with distinct character (strongly localized $\sigma$ orbitals and less localized $\pi$-orbitals in the charge-spread sense) rather than averaging them together. . 
 Any unitary rotation that mixes orbitals with distinct character necessarily spreads electronic population over multiple atomic centers, thereby reducing orbital locality and being variationally disfavored.

\section{Applying random perturbation of to the initial occupied orbitals before localization.} \label{sec:URot}

For N orbitals to rotate by $\theta^\circ$, we make an anti-symmetric $(N,N)$ matrix $\mat A$, in which each element of the strictly upper-triangle (i.e., $\mat A[i,j]$ with $j < i$) is drawn from a normal distribution with mean zero and standard deviation  $\dfrac{2 \pi \,\theta ^\circ}{180 ^\circ}$ (given in radians):

Note: the set parameter for the random rotation angle applies for each matrix element, and therefore each rotation between a unique pair of occupied orbitals separately. 
This choice here is size consistent and has correct scaling in terms of hypothetical non-interacting sub-systems of a larger system.

The unitary rotation matrix $\mat U$ is found as the matrix exponential:
\begin{align}
  \mat U = e^{\mat A}
\end{align}
As $\mat A$ is real and antisymmetric (and therefore also anti-hermitian), $\mat U$ is not only unitary but also orthonormal.
The unitarily transformed occupied orbital matrix $\tilde{\mat C}$ is given by:
 \begin{align}
\tilde{\mat C}=\mat C \; \mat U
\end{align}

\section{On different forms of the FB localization functional} \label{sec:diffBoys}
Here we explain why the \Eqref{eq:LocFnBoys1} and \Eqref{eq:LocFnBoys2} are equivalent. 
Let $\phi_i$ functions form an orthonormal basis set.
To simplify, let us show that in a one-dimensional case:
\begin{align}
&\braket{\phi_{i}\phi_{i}| \left(x_1-x_2 \right)^2|\phi_{i}\phi_{i}}= \notag \\
&=\int \int \phi^*_i(x_1)\phi^*_i(x_2) (x_1-x_2)^2  \phi_i(x_1)\phi_i(x_2) \mathrm{d}x_1\mathrm{d}x_2= \notag \\
&=\int \int \phi^*_i(x_1)\phi^*_i(x_2) (x^2_1-2x_1x_2+x^2_2)  \phi_i(x_1)\phi_i(x_2) \mathrm{d}x_1\mathrm{d}x_2 =\notag \\
&=\int \int \phi^*_i(x_1)\phi^*_i(x_2) x_1^2  \phi_i(x_1)\phi_i(x_2) \mathrm{d}x_1\mathrm{d}x_2 \notag \\ 
&\qquad-2\int \int \phi^*_i(x_1)\phi^*_i(x_2) x_1 x_2  \phi_i(x_1)\phi_i(x_2) \mathrm{d}x_1\mathrm{d}x_2  \notag \\
&\qquad+\int \int \phi^*_i(x_1)\phi^*_i(x_2) x_2^2  \phi_i(x_1)\phi_i(x_2) \mathrm{d}x_1\mathrm{d}x_2=  \notag \\
&=\int \phi^*_i(x_1) x_1^2  \phi_i(x_1) \mathrm{d}x_1   \int  \phi^*_i(x_2) \phi_i(x_2)  \mathrm{d}x_2\notag \\
&\qquad-2\int \phi^*_i(x_1) x_1 \phi_i(x_1) \mathrm{d}x_1   \int  \phi^*_i(x_2)x_2  \phi_i(x_2)  \mathrm{d}x_2\notag \\
&\qquad+\int \phi^*_i(x_1)  \phi_i(x_1) \mathrm{d}x_1   \int  \phi^*_i(x_2) x_2^2  \phi_i(x_2)  \mathrm{d}x_2\notag \\
&=\braket{\phi_{i}| x^2 |\phi_{i}}-2\braket{\phi_{i}| x |\phi_{i}}\braket{\phi_{i}| x |\phi_{i}}+\braket{\phi_{i}| x^2 |\phi_{i}}=\notag \\
&=2\braket{\phi_{i}| x^2 |\phi_{i}}-2 \braket{\phi_{i}| x |\phi_{i}} ^2 
\end{align}
We can generalize this result to a three-dimensional case. As 
\begin{align}
\left|\hat{\mathbf{ r}}_1-\hat{\mathbf{ r}}_2 \right|^2=\left(x_1-x_2 \right)^2+\left(y_1-y_2 \right)^2+\left(z_1-z_2 \right)^2,
\end{align}
it is not difficut to show that
\begin{align}
 \sum_{i=1}^{N_\mathrm{orb}} \braket{\phi_{i}\phi_{i}| \left|\hat{\mathbf{ r}}_1-\hat{\mathbf{ r}}_2 \right|^2|\phi_{i}\phi_{i}}=
2 \sum_{i=1}^{N_\mathrm{orb}}  \left (\braket{\phi_{i}| \hat{\mathbf{ r}}^2| \phi_{i}} - \braket{\phi_{i}| \hat{\mathbf{ r}}| \phi_{i}}^2 \right )
\end{align}

So minimizing the localization functional $ \sum_{i=1}^{N_\mathrm{orb}} \braket{\phi_{i}\phi_{i}| \left|\hat{\mathbf{ r}}_1-\hat{\mathbf{ r}}_2 \right|^2|\phi_{i}\phi_{i}}$ is equivalent to minimizing the localization functional $\sum_{i=1}^{N_\mathrm{orb}}  \left (\braket{\phi_{i}| \hat{\mathbf{ r}}^2| \phi_{i}} - \braket{\phi_{i}| \hat{\mathbf{ r}}| \phi_{i}}^2 \right )$.\\

\end{document}